\documentclass[11pt]{article}

\usepackage[margin=1in]{geometry}
\usepackage{amsmath, amssymb}
\usepackage{graphicx}
\usepackage{hyperref}
\usepackage{caption}
\usepackage{subcaption}
\usepackage{booktabs}
\usepackage{lmodern}
\usepackage[T1]{fontenc}  
\usepackage{float}  
\usepackage[english]{babel}

\title{Causal and Predictive Modeling of Short-Horizon Market Risk and Systematic Alpha Generation Using Hybrid Machine Learning Ensembles}
\author{Aryan Ranjan\\
University of Oxford\\
United Kingdom\\
\texttt{aryan.ranjan@stcatz.ox.ac.uk}}

\date{\today}

\begin{document}

\emergencystretch=2em

\maketitle

\begin{abstract}
\noindent
We present a systematic trading framework that forecasts short-horizon market risk, identifies its underlying drivers, and generates alpha using a hybrid machine learning ensemble built to trade on the resulting signal. The framework integrates neural networks with tree-based voting models to predict five-day drawdowns in the S\&P 500 ETF (SPY), leveraging a cross-asset feature set spanning equities, fixed income, foreign exchange, commodities, and volatility markets. Interpretable feature attribution methods reveal the key macroeconomic and microstructural factors that differentiate high-risk (crash) from benign (non-crash) weekly regimes. Empirical results show a Sharpe ratio of $2.51$ and an annualized CAPM alpha of $+0.28$, with a market beta of $0.51$, indicating that the model delivers substantial systematic alpha with limited directional exposure during the $2005–2025$ backtest period. Overall, the findings underscore the effectiveness of hybrid ensemble architectures in capturing nonlinear risk dynamics and identifying interpretable, potentially causal drivers, providing a robust blueprint for machine learning–driven alpha generation in systematic trading.
\end{abstract}

\section{Introduction}
Recent equity market dynamics have exhibited short-term, irregular deviations and pronounced volatility clustering, often temporarily decoupling from established macro-financial drivers. Extreme market dislocations—such as the rapid deposit flight during the March 2023 regional bank failures, the spike in implied and realized volatility following President Trump’s \textit{Liberation Day} tariffs in April 2025, and the persistent, sharp weekly drawdowns from all-time highs in 2024–2025—demonstrate the limited predictive power of conventional risk models in anticipating short-horizon shocks. As a result, the identification and forecasting of near-term market risk regimes has become a critical prerequisite for systematic alpha generation and adaptive portfolio risk management. 

Traditional econometric frameworks—such as Generalized Autoregressive Conditional Heteroskedasticity (GARCH) and linear factor models—are fundamentally limited in capturing the highly nonlinear and cross-asset dependencies that increasingly govern intramonth market behavior. The necessity of modeling high-dimensional feature spaces and complex interaction effects motivates the deployment of modern machine learning techniques. Although prior hybrid approaches have primarily combined GARCH with Long Short-Term Memory (LSTM) networks, the significant computational burden imposed by LSTMs—due to their complex gating structures and large parameter space—limits their practicality in latency-sensitive financial forecasting applications. Instead, we posit that a hybrid architecture combining neural networks (NNs) and tree-based ensembles offers a particularly compelling solution. The deep neural layers are engineered to capture complex, non-linear mappings and latent state representations from multi-modal input features, while the tree ensembles (e.g., Extreme Gradient Boosted Trees, CatBoost) provide robust handling of sparse, non-Gaussian feature distributions and emphasize interpretable feature splits and threshold effects on large, noisy datasets. 

This study develops a predictive system that leverages such hybrid architectures to generate actionable signals for five-day SPY drawdowns, utilizing this metric as a granular proxy for short-horizon systematic risk. We undertake extensive feature engineering, integrating data streams across five primary financial domains: equities, bonds, foreign exchange, commodities, and volatility products. The ensemble model is then trained on this feature-rich representation of the global market state to generate directional signals. Crucially, we position this effort not as a pursuit of pure probabilistic prediction but as a framework for systematic signal and alpha generation, where the capacity to produce persistent, risk-adjusted excess returns is rigorously grounded by causal inference analysis for true interpretability of weekly risk classifications (crash vs. non-crash regime). 

Our results demonstrate that a hybrid neural–tree ensemble possesses statistically significant predictive power in identifying latent risk regimes and anticipating short-horizon corrections. This methodology successfully bridges the gap between traditional, interpretable macro-driven trading signals and modern, data-driven machine learning, providing both academic insight into non-linear market causality and practical utility for constructing robust, adaptive systematic trading strategies.

\section{Methodology}
\subsection{Pipeline Flow}
We split this work into two different phases: (1) predictive signal and (2) quintile-based strategy to trade on the signal systematically. The overall pipeline flow for each phase is illustrated below: 
\[
\underbrace{\mathbf{X}_{t-\tau:t}}_{\text{Selected features}}
\;\xrightarrow[\text{MLP + Tree Ensemble}]{\text{Model Prediction}}\;
\underbrace{\mathbf{Z}_t}_{\text{Pred. risk signal}}
\;\xrightarrow[\text{Rolling/Temporal Aggregation}]{\text{ Strategy Signal}}\;
\underbrace{s_t}_{\text{Strategy Signal}}
\]

\[
s_t \;\xrightarrow[\text{Quintile Mapping}]{\text{Signal → Position}} 
p_t = f(s_t) \in \{-1, 1\} 
\;\xrightarrow[\text{Portfolio Construction}]{\text{Apply Positions}}\; 
R_t^{\text{strategy}} 
\;\xrightarrow[\text{Performance Metrics}]{\text{Evaluation}}\; 
\text{Risk Score}
\]
\subsection{Data and Features}
\paragraph{Investment Universe.}  
We construct our predictive framework using a broad cross-section of market instruments to capture equity, fixed income, volatility, commodity, and currency dynamics. Specifically, our universe $\mathcal{U}$ consists of:

\begin{table}[H]
\centering
\caption{Investment Universe $\mathcal{U}$ used for feature engineering.}
\label{tab:universe}
\begin{tabular}{ll}
\toprule
Asset Class & Symbols / Proxies \\
\midrule
Equities & SPY, QQQ, IWM, TLT \\
Volatility & VIX \\
Commodities & GLD, CL=F \\
Foreign Exchange & DX-Y.NYB, EURUSD=X, JPYUSD=X \\
Treasuries & TNX, IRX \\
\bottomrule
\end{tabular}
\end{table}

We denote the historical adjusted price series of each asset $i \in \mathcal{U}$ at time $t$ as $P_{i,t}$, and compute logarithmic returns:
\begin{equation}
r_{i,t} = \ln(P_{i,t}) - \ln(P_{i,t-1}).
\end{equation}

\paragraph{Target Variable.}  
The primary objective of our model is to predict the probability of a significant market drawdown for the SPY ETF over a short-term horizon. Formally, we define the binary target $y_t$ as:
\begin{equation}
y_t = \mathbf{1}\Bigg\{ \sum_{k=1}^{h} r_{\text{SPY}, t+k} \leq -\delta \Bigg\},
\end{equation}
where $h=5$ trading days, and $\delta = 1\%$. This definition captures extreme short-term downside risk over the next trading week.

\paragraph{Feature Engineering.}  
Our feature space $\mathbf{X}_t$ is designed to capture both statistical characteristics of returns and economically motivated signals. We engineer features along multiple dimensions:

\begin{enumerate}
    \item \textbf{Time-Series Moments (Statistical Features):}  
    For each asset $i$, we compute rolling window statistics over multiple horizons $w \in \{21, 63\}$ days to capture short- and medium-term dynamics:
    \begin{align}
        \text{Volatility: } & \sigma_{i,t}^{(w)} = \text{std}(r_{i,t-w+1:t}) \\
        \text{Skewness: } & \gamma_{i,t}^{(w)} = \frac{\frac{1}{w}\sum_{s=0}^{w-1}(r_{i,t-s}-\bar{r}_{i,t}^{(w)})^3}{(\sigma_{i,t}^{(w)})^3} \\
        \text{Kurtosis: } & \kappa_{i,t}^{(w)} = \frac{\frac{1}{w}\sum_{s=0}^{w-1}(r_{i,t-s}-\bar{r}_{i,t}^{(w)})^4}{(\sigma_{i,t}^{(w)})^4} - 3 \\
        \text{Shannon Entropy: } & H_{i,t}^{(w)} = - \sum_{j=1}^{B} p_j \log p_j
    \end{align}
    where $p_j$ are normalized histogram counts over $B=30$ bins. These features capture volatility, asymmetry, tail behavior, and uncertainty in return distributions.

    \item \textbf{Hurst Exponent (Persistence/Mean-Reversion Indicator):}  
    We estimate the Hurst exponent $H_{i,t}^{(\tau)}$ over multiple scales $\tau \in \{\text{short}=16, \text{medium}=64, \text{long}=256\}$ using a vectorized rescaled range methodology centered on powers of two for numerical stability. This quantifies the degree of long-term memory in the price series.

    \item \textbf{Cross-Asset Relations:}  
    To capture dependencies between SPY and other assets:
    \begin{align}
        \beta_{i,\text{SPY},t}^{(w)} &= \text{RollingOLS}\big(r_i \sim r_\text{SPY}, w \big) \\
        \rho_{i,\text{SPY},t}^{(w)} &= \text{Corr}(r_{i,t-w+1:t}, r_{\text{SPY},t-w+1:t})
    \end{align}

    \item \textbf{Information-Theoretic Measures:}  
    Rolling Kullback-Leibler divergence quantifies shifts in return distributions relative to longer reference windows (21 days for short term, 126 days for long term):
    \begin{equation}
        \text{KL}_{i,t}^{(w_\text{curr}, w_\text{ref})} = \sum_{j=1}^{B} p_j^\text{curr} \log \frac{p_j^\text{curr}}{p_j^\text{ref}}
    \end{equation}

\end{enumerate}

\paragraph{Feature Construction Summary.}  
The resulting feature space $\mathbf{X}_t$ contains 178 engineered variables, spanning 5423 total observations over the 2005-2025 period. These features are carefully forward-filled and standardized to handle missing values and scale disparities.

\begin{figure}[h]
    \centering
    \includegraphics[width=\textwidth]{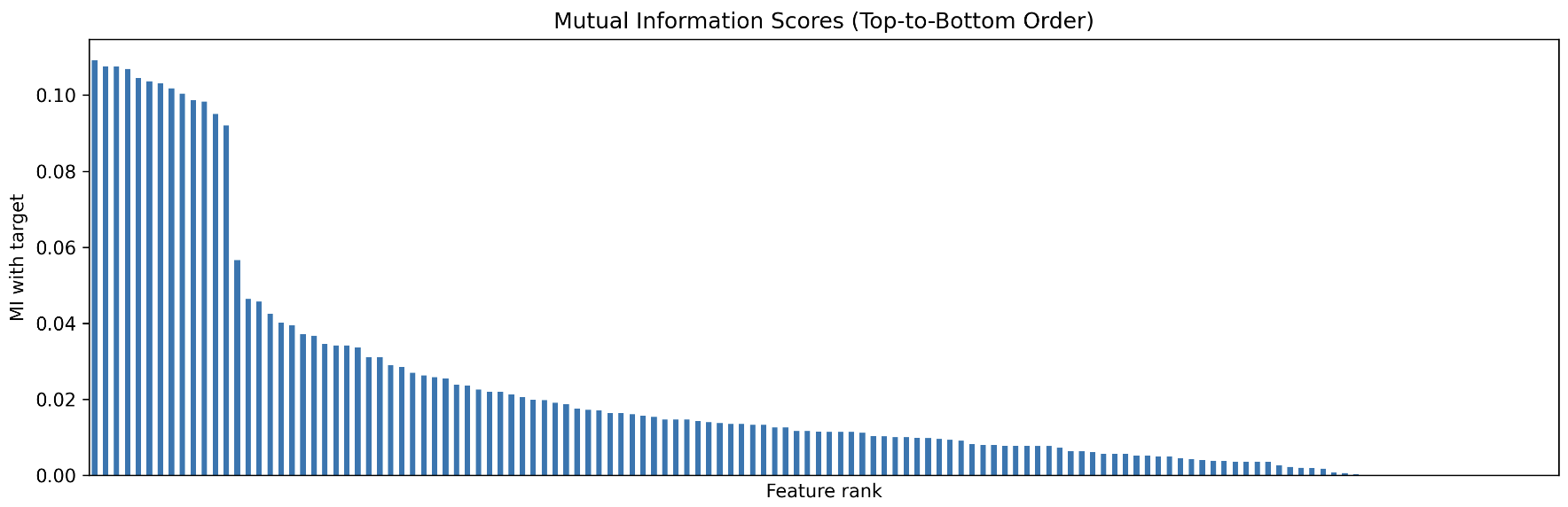}
    \caption{Mutual information scores for all candidate features.}
    \label{fig:mutual_info}
\end{figure}

\paragraph{Feature Selection via Mutual Information.}  
Given the high-dimensional feature space, we apply an initial statistical filtration to reduce redundancy and remove low-informative variables. Specifically, 
\begin{itemize}
    \item \textbf{Low variance removal}: Features with near-zero variance ($<10^{-4}$) are removed because they carry little to no predictive information.
    \item \textbf{High-correlation removal}: We remove one variable from a pair of highly Pearson correlated features ($ \geq 0.95$) to avoid multicollinearity issues. 
\end{itemize}
This initial filtration narrows our feature space down to 134 filters. From here, we use mutual information (MI) to rank features based on their predictive relevance with respect to the target $y_t$:
\begin{equation}
\text{MI}(X_j, y) = \sum_{x \in X_j} \sum_{y \in \{0,1\}} p(x,y) \log \frac{p(x,y)}{p(x)p(y)}
\end{equation}
After computing MI scores for all candidate features, we select the top $80$ features that exhibit the strongest dependency with the target. Figure~\ref{fig:mutual_info} illustrates the MI distribution across the full candidate set.

From this graph, there is a steep drop-off in MI scores after the 13th feature. The top features are as follows: 

\begin{table}[H]
\centering
\caption{Top 13 Features by Mutual Information Score}
\label{tab:top13_features_mi}
\begin{tabular}{l c}
\toprule
\textbf{Feature} & \textbf{Mutual Information Score} \\
\midrule
\texttt{CL=F\_hurst\_short} & 0.109211 \\
\texttt{JPYUSD=X\_hurst\_short} & 0.107575 \\
\texttt{IRX\_hurst\_short} & 0.107422 \\
\texttt{GLD\_hurst\_short} & 0.106937 \\
\texttt{TNX\_hurst\_short}  & 0.104463 \\
\texttt{DX-Y.NYB\_hurst\_short} & 0.103586 \\
\texttt{EURUSD=X\_hurst\_short} & 0.103037 \\
\texttt{TLT\_hurst\_short} & 0.101677 \\
\texttt{VIX\_hurst\_short} & 0.100433 \\
\texttt{QQQ\_hurst\_short} & 0.098766 \\
\texttt{IWM\_hurst\_short} & 0.098508 \\
\texttt{IRX\_vol\_63d} & 0.095007 \\
\texttt{SPY\_hurst\_short} & 0.092049 \\
\bottomrule
\end{tabular}
\end{table}
\paragraph{Feature Insights}
\label{sec: feature_insights}
A striking insight from the MI analysis is that short-horizon Hurst exponents across multiple asset classes exhibit the highest statistical dependence with 5-day SPY drawdowns, with only one medium-term volatility measure showing comparable dependence. This highlights two important aspects. First, the short time horizon captured by the Hurst exponents suggests that \emph{microstructure-level persistence or mean-reversion patterns} contain meaningful information for anticipating short-term drawdowns. These dynamics likely encode subtle regime shifts that standard longer-horizon volatility or correlation measures would miss.

Second, and perhaps more surprisingly, the features with the strongest MI are largely drawn from \emph{cross-asset classes—commodities, FX, and Treasuries—rather than SPY itself or other equity indices}. Intuitively, one might expect SPY returns or other equity indices to dominate as predictors, especially during periods of strong upward trends or record highs. Yet, our MI analysis reveals that signals from oil prices, currencies, and interest rates show greater statistical association with short-term SPY drawdowns than SPY itself. This implies that the equity market is often \emph{reactive} rather than \emph{proactive}: shocks in global macro or financial markets tend to precede equity declines, with equity indices following rather than leading.

This provides a valuable lens for interpreting current AI-related equity hype and bubble concerns. Even as headline indices surge on AI enthusiasm, the cross-asset signals embedded in our top features suggest that latent macroeconomic pressures—shifts in oil, currency, or interest rate dynamics—often appear \emph{before} equities decline. Put differently, while investors focus on soaring equity prices, important early warning signs \emph{may already exist elsewhere}. This shows why monitoring multiple asset classes outside of the equity universe can give a better early signal of potential drawdowns. 

\paragraph{Final Feature Selection} To ensure that the model has access to potentially relevant signals beyond the most obvious ones, we retain the top 80 features rather than truncating at the steep drop-off after the 13th feature in Figure~\ref{fig:mutual_info}. There are several reasons for this choice:
\begin{itemize}
    \item \textbf{Non-linear interactions}: MI captures individual feature-target dependency, but does not account for interactions between features. Features with low marginal MI may still contribute meaningfully in combination with other features through non-linear relationships captured by tree-based models or neural networks.
    \item \textbf{SHAP explainability}: For interpretability via SHapley Additive exPlanations (SHAP), it is beneficial to include a broader set of features. Weak predictors may help the model refine contributions of stronger features and reveal subtle dependencies in the market dynamics.
    \item \textbf{Ensemble diversity}: The ensemble incorporates multiple model types (MLP, RF, XGBoost, CatBoost). Some models can extract signal from weaker or correlated features, which would otherwise be ignored if we limited ourselves to the top 13 features.
\end{itemize}
This rigorous filtering ensures that the model is trained on features with the highest signal-to-noise ratio, combining both statistical and economically motivated signals while mitigating redundancy.

\subsection{Ensemble Models}
To leverage complementary strengths of different model classes, we construct a \textbf{soft-voting ensemble} composed of neural networks and tree-based gradient boosting models. Let the \textbf{base learners} be denoted as
\[
\mathcal{M} = \{ M_1, M_2, \dots, M_K \},
\]
where \(M_1\) is a shallow \textbf{Multi-Layer Perceptron (MLP)} and \(M_2, \dots, M_K\) are gradient-boosted decision trees (XGBoost, CatBoost).
\subsubsection{Base Learners}

\paragraph{Multi-Layer Perceptron (MLP)}
The MLP is designed to capture temporal and non-linear interactions in the engineered feature set. For an input vector \(\mathbf{x} \in \mathbb{R}^d\), the MLP computes:
\[
\mathbf{h}^{(1)} = \sigma \big( \mathbf{W}^{(1)} \mathbf{x} + \mathbf{b}^{(1)} \big), \quad
\mathbf{h}^{(l)} = \sigma \big( \mathbf{W}^{(l)} \mathbf{h}^{(l-1)} + \mathbf{b}^{(l)} \big), \quad
\hat{y} = \text{softmax}(\mathbf{W}^{(L)} \mathbf{h}^{(L-1)} + \mathbf{b}^{(L)}),
\]
where \(L\) is the number of layers, \(\sigma(\cdot)\) is an elementwise activation function (ReLU or tanh), and \((\mathbf{W}^{(l)}, \mathbf{b}^{(l)})\) are trainable weights and biases.

We intentionally hardcoded a shallow MLP architecture (1-3 hidden layers, moderate width) because our dataset, though rich in features, is limited in effective sample size (5,000--6,000 daily observations after filtering). A shallow network mitigates overfitting while still capturing temporal dependencies.

\paragraph{Gradient Boosted Trees (XGBoost, CatBoost)}
For structured tabular features, we use gradient boosting classifiers. A single tree \(T_m(\mathbf{x})\) outputs a prediction for the log-odds of a binary target:
\[
f_m(\mathbf{x}) = \sum_{j=1}^{J_m} w_j \mathbb{I}\{\mathbf{x} \in R_j\},
\]
where \(R_j\) are terminal regions (leaves) and \(w_j\) are leaf weights learned via gradient descent on the logistic loss:
\[
\mathcal{L} = - \sum_{i=1}^N \left[ y_i \log \hat{p}_i + (1-y_i) \log (1-\hat{p}_i) \right].
\]

XGBoost and CatBoost sequentially fit \(M\) trees to residuals of the previous model, producing an additive ensemble:
\[
F(\mathbf{x}) = \sum_{m=1}^M T_m(\mathbf{x}).
\]

CatBoost further handles categorical features and ordering bias, while XGBoost is optimized for speed and regularization.

\subsubsection{Hyperparameter Optimization}
Each base learner undergoes grid search with TimeSeriesSplit cross-validation to maximize ROC-AUC. Formally, for model \(M_k\) and hyperparameter set \(\Theta_k\), we solve:
\[
\hat{\theta}_k = \arg\max_{\theta \in \Theta_k} \frac{1}{S} \sum_{s=1}^S \text{AUC}\big(y^{(s)}, M_k^{(\theta)}(\mathbf{X}^{(s)})\big),
\]
where \(S\) is the number of folds and \((\mathbf{X}^{(s)}, y^{(s)})\) are the train-test splits respecting temporal order.

\subsubsection{Soft Voting Ensemble}
The final ensemble aggregates base learner probabilities using soft voting:
\[
\hat{p}(\mathbf{x}) = \frac{1}{K} \sum_{k=1}^K \hat{p}_k(\mathbf{x}),
\]
where \(\hat{p}_k(\mathbf{x})\) is the predicted probability of a 5-day SPY drawdown \(\ge 1\%\) from base learner \(M_k\). The binary prediction is then:
\[
\hat{y} = \mathbb{I}\{ \hat{p}(\mathbf{x}) \ge 0.5 \}.
\]

\noindent
This ensemble approach leverages complementary strengths:
\begin{itemize}
    \item The MLP captures \emph{non-linear temporal dynamics} and interactions across features.
    \item Gradient-boosted trees model \emph{structured non-linearities and feature importance}, providing interpretability via SHAP.
    \item Voting stabilizes predictions, \emph{reducing variance} of individual learners and improving generalization.
\end{itemize}

By combining these learners mathematically, the model can capture both subtle temporal signals and robust tabular patterns in equity market dynamics, which is critical for generating accurate risk signals in financial time series.

\section{Results}
\subsection{Model Performance}

We evaluate the predictive performance of our ensemble model on the full feature set. While hyperparameter tuning employed rolling time-series cross-validation, the final ensemble is trained on the entire dataset. Metrics such as ROC-AUC, precision, and recall therefore reflect in-sample performance.

\begin{table}[h]
\centering
\caption{Validation Set Classification Report}
\label{tab:classification_report}
\begin{tabular}{lcccc}
\toprule
\textbf{Class} & \textbf{Precision} & \textbf{Recall} & \textbf{F1-score} & \textbf{Support}  \\ 
\midrule
0 & 0.95 & 0.99 & 0.97 & 4232 \\
1 & 0.95 & 0.82 & 0.88 & 1191 \\
\midrule
\textbf{Accuracy}      & & & 0.95 & 5423 \\
\textbf{Macro Avg}     & 0.95 & 0.90 & 0.92 & 5423 \\
\textbf{Weighted Avg}  & 0.95 & 0.95 & 0.95 & 5423 \\
\bottomrule
\end{tabular}
\end{table}

\begin{figure}[h]
\centering
\includegraphics[width=0.6\textwidth]{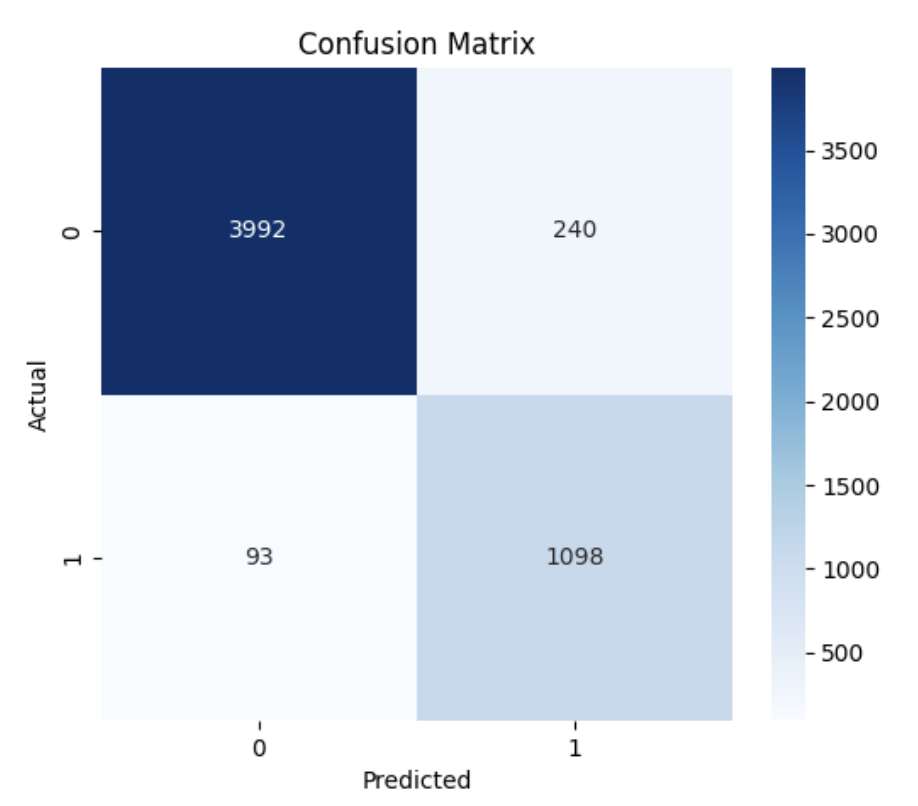}
\caption{Confusion matrix on the validation set. True non-crash weeks are in the first row/column, true crash weeks in the second.}
\label{fig:confusion_matrix}
\end{figure}

Non-crash detection is extremely robust. Class 0 recall of 0.99 indicates that the model almost never mislabels normal weeks as crashes, minimizing false alarms—a critical property for any risk-adjusted strategy. Crash detection is also strong but selective. Class 1 recall of 0.82 shows the ensemble correctly identifies the majority of high-risk weeks while maintaining high precision (0.95). This reflects a trade-off aligned with risk management: avoiding overreacting to noise while still capturing genuine SPY drawdowns. Weighted averages show overall high accuracy (0.95) and F1-score (0.95), suggesting the model captures the cross-asset and temporal patterns leading to SPY drawdowns without overfitting to either class.

From Figure~\ref{fig:confusion_matrix}, we see that most misclassifications (93 false negatives) correspond to crash weeks not flagged by the model. Examining the corresponding dates reveals that these periods are typically mild drawdowns that, while exceeding the 1\% threshold, do not exhibit systemic signals captured by the ensemble (e.g., low volatility clustering, muted cross-asset correlation). Conversely, the 240 false positives often occur in weeks with transient market turbulence that did not materialize into a 5-day SPY drop.

Overall, the model’s predictive output is consistent with a practical risk-management approach: it emphasizes precision over recall for crashes, ensuring that actionable high-risk signals are credible, while normal weeks are rarely over-flagged. This aligns with how a quantitative trading desk would leverage risk signals: favoring reliability of alerts over exhaustive detection.

\subsection{Feature Attribution and Causal Insights}

\begin{figure}[H]
    \centering
        \centering
        \includegraphics[width=\textwidth]{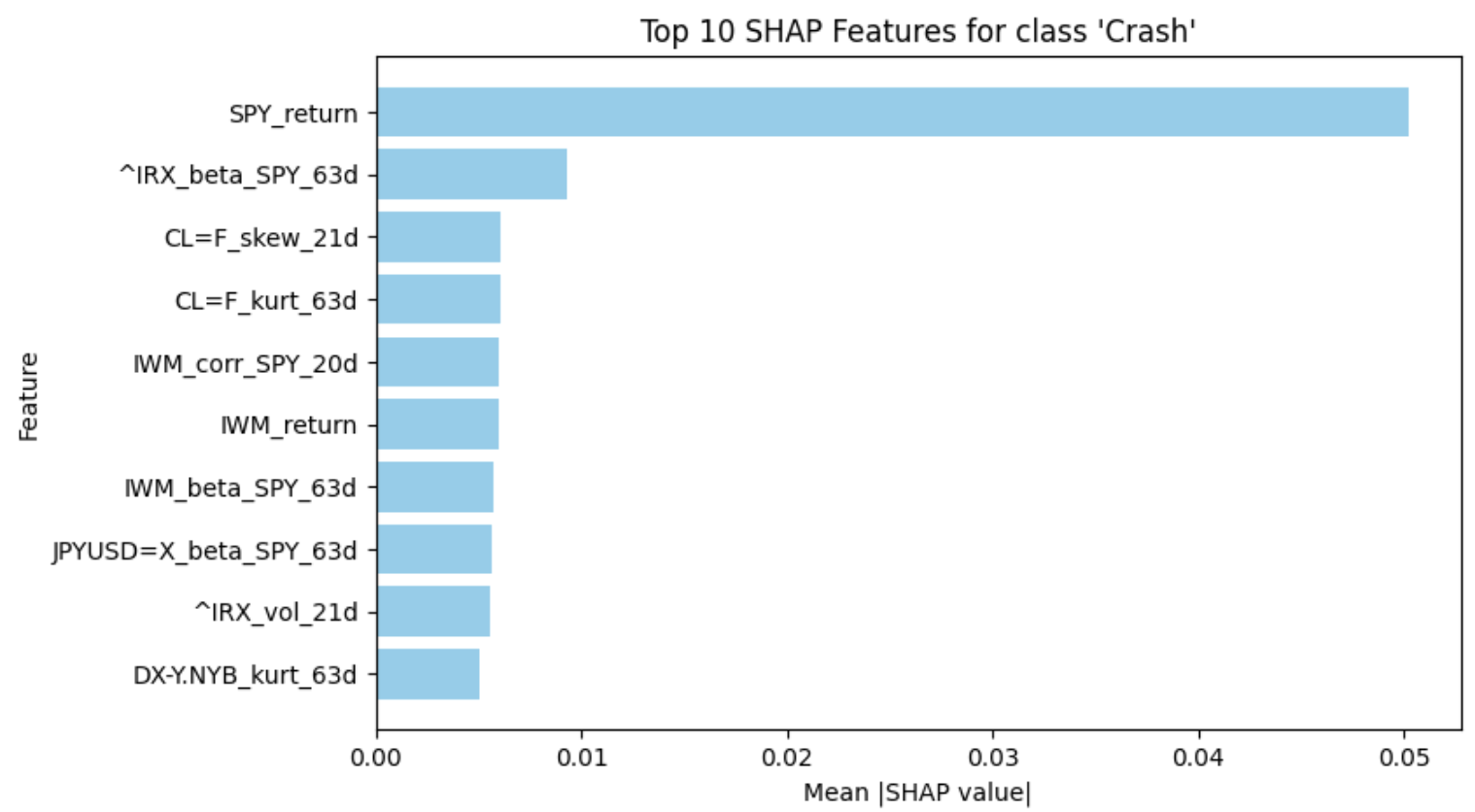}
        \caption{Crash weeks}
        \label{fig:shap_crash}
\end{figure}

\begin{figure}[H]
    \centering
    \includegraphics[width=\textwidth]{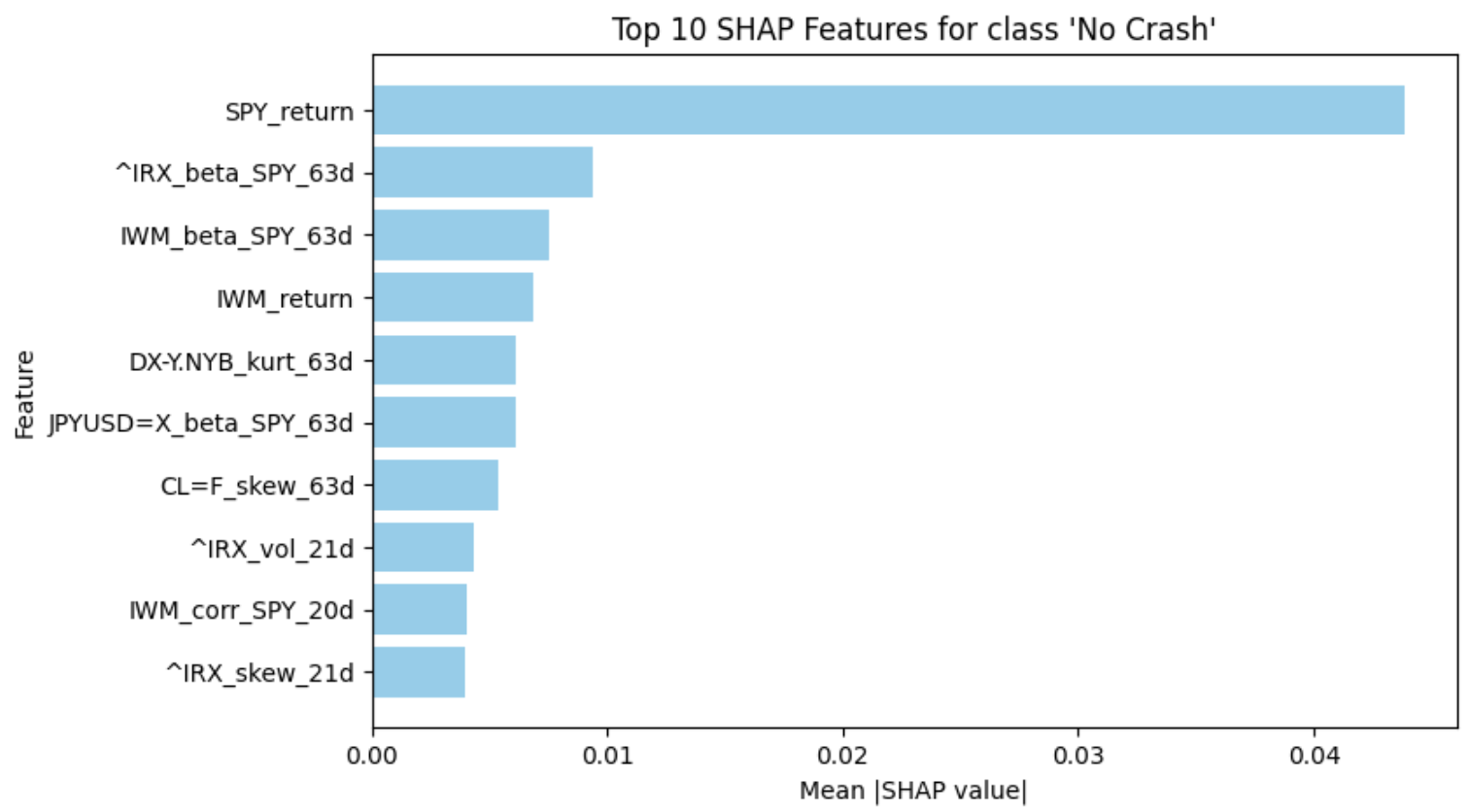}
    \caption{Non-crash weeks}
    \label{fig:shap_non_crash}
\end{figure}

To interpret the drivers of crash risk, we apply SHAP to the ensemble model. Figures~\ref{fig:shap_crash} and~\ref{fig:shap_non_crash} show the SHAP summary plots for crash and non-crash weeks, respectively.

Interestingly, while MI analysis highlighted Hurst exponents across assets as among the strongest predictors, SHAP shows that these features play a minimal role in the model’s final predictions. This divergence arises because MI measures \emph{unconditional dependence} between a feature and the target, capturing general market fragility signals such as trending or mean-reverting regimes. In contrast, SHAP values reflect \emph{conditional contributions within the full feature set.} Once short-term, cross-asset signals are included, the Hurst metrics’ predictive power is largely absorbed by more immediate drivers. Economically, this suggests that while long-term market autocorrelation sets the stage for fragility, the timing of crashes is dominated by short-term shocks and cross-asset interactions.

During crash weeks, the model assigns significant weight to cross-asset measures, particularly in commodities, FX, and Treasuries. Features such as SPY returns, 63-day rolling beta of SPY returns with respect to the 3M Treasury Yield, oil higher-order moments (skew and kurtosis), and IWM returns and beta emerge as influential. These findings can be economically interpreted as follows:

\begin{itemize}
    \item \textbf{SPY returns:} While MI suggested SPY returns alone are weakly predictive, SHAP reveals their conditional importance. SPY drawdowns amplify crash risk in the presence of systemic cross-asset stress, acting as a signal of broader market fragility.
    
    \item \textbf{Interest rate sensitivity (IRX beta with SPY):} Short-term rate movements interacting with equities signal changing monetary conditions. Rising rates or increased rate-equity correlation during crash weeks heighten systemic stress, consistent with historical episodes where monetary tightening contributed to equity drawdowns.
    
    \item \textbf{Oil skew and kurtosis:} Tail measures of oil prices capture the probability of extreme directional moves. Commodity extremes often coincide with market-wide risk-off events, making them leading indicators of macro-financial stress.
    
    \item \textbf{IWM returns and beta:} Small-cap equities are more sensitive to liquidity shocks and shifts in risk appetite than large-cap tech-heavy indices (e.g., QQQ). Their outsized contribution reflects the model’s detection of systemic market vulnerability beyond headline tech risk. Although QQQ and large-cap tech dominate headlines, especially during recent AI-driven market cycles, the 2005–2025 window includes multiple crises (e.g., 2008–09 GFC, 2011 Eurozone stress, 2020 COVID drawdown) where small- and mid-cap equities historically led or amplified market stress. Their outsized SHAP contribution highlights that \emph{systemic vulnerability often manifests first in smaller, more liquidity-sensitive segments rather than the tech-heavy Nasdaq in this period of study}.
\end{itemize}

For non-crash weeks, the SHAP analysis identifies a partially overlapping but economically distinct set of influential features: SPY and IWM returns and beta, Treasury sensitivity (IRX beta), Dollar Index kurtosis, JPY/USD beta, and oil return skewness. Several key contrasts emerge compared with crash weeks:

\begin{itemize}
    \item \textbf{SPY and IWM returns:} Still contribute meaningfully, but generally reduce predicted crash probability. Positive or stable equity performance across broad- and small-cap indices provides conditional reassurance against imminent market stress.
    \item \textbf{Macro tail risk indicators:} Crash weeks are dominated by commodity stress signals, but \emph{non-crash weeks emphasize FX market indicators}. Stable or trending FX dynamics contribute to risk dampening rather than amplification. Thus, the FX market acts as a risk redistribution mechanism: orderly currency moves reflect healthy cross-border capital flows and monetary stability, stabilizing equities even when minor volatility arises elsewhere.  
    \item \textbf{Cross-asset interactions and regime asymmetry:} Non-crash weeks reflect stable, trending dynamics, whereas crash weeks are triggered by extreme deviations in oil, Treasury yields, or equity returns. This asymmetry suggests the model captures the \emph{functional role of different asset classes under stress vs. calm regimes}: commodities are key risk triggers, while FX provides stabilizing signals.  
\end{itemize}

The SHAP analysis reveals a clear divergence between crash and non-crash regimes, highlighting the asymmetric roles of different asset classes in systemic risk propagation. During crashes, commodities and small-cap equities act as \emph{shock amplifiers}, signaling liquidity stress and macro-financial fragility, whereas SPY returns act as a conditional amplifier in the context of these broader signals. In contrast, non-crash weeks show that FX and Treasury sensitivities provide \emph{stabilizing feedback}, reflecting orderly capital flows and well-functioning risk redistribution mechanisms. Economically, this suggests that systemic vulnerability is not just a function of asset volatility, but of \emph{cross-asset interactions and regime-dependent transmission channels}. Long-term fragility indicators, such as Hurst exponents, set the backdrop for market susceptibility, but the realized timing and magnitude of crashes are driven by short-term shocks that propagate unevenly across assets. This nuanced understanding can guide both predictive modeling and macro-financial risk monitoring, emphasizing that effective early-warning signals require attention to both which assets are stressed and how stress interacts across the financial ecosystem. 

\subsection{Risk Quantile Analysis}

\begin{figure}[H]
\centering
\includegraphics[width=\textwidth]{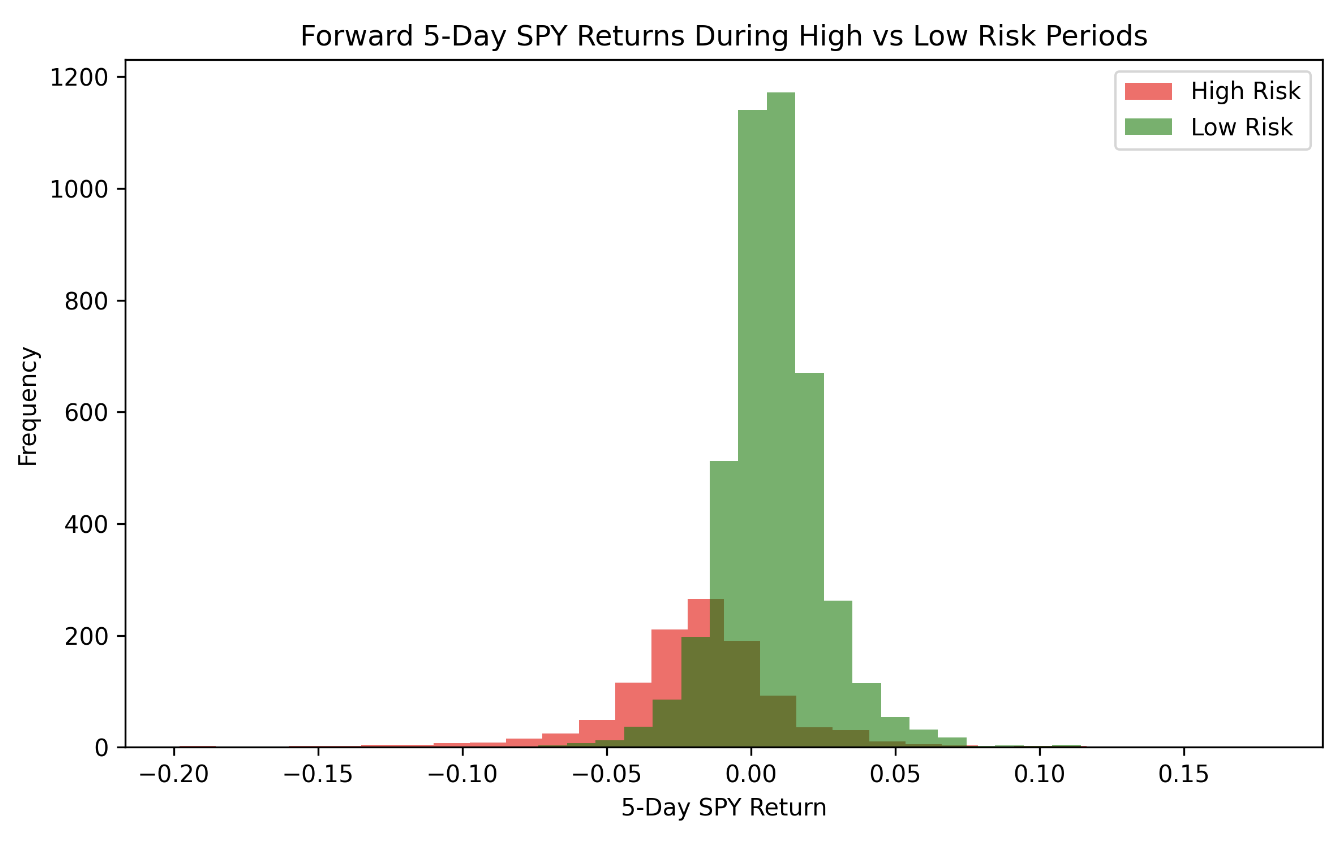}
\caption{Cumulative SPY returns stratified by LOW-risk and HIGH-risk periods.}
\label{fig:dist}
\end{figure}

To translate model predictions into actionable signals, we rank all weeks by predicted crash probability and segment them into quintiles. Figures~\ref{fig:risk_quantile_plot} and~\ref{fig:dist} illustrate the realized SPY returns conditional on predicted risk levels. Several key insights emerge:

\begin{figure}[H]
\centering
\includegraphics[width=0.9\textwidth]{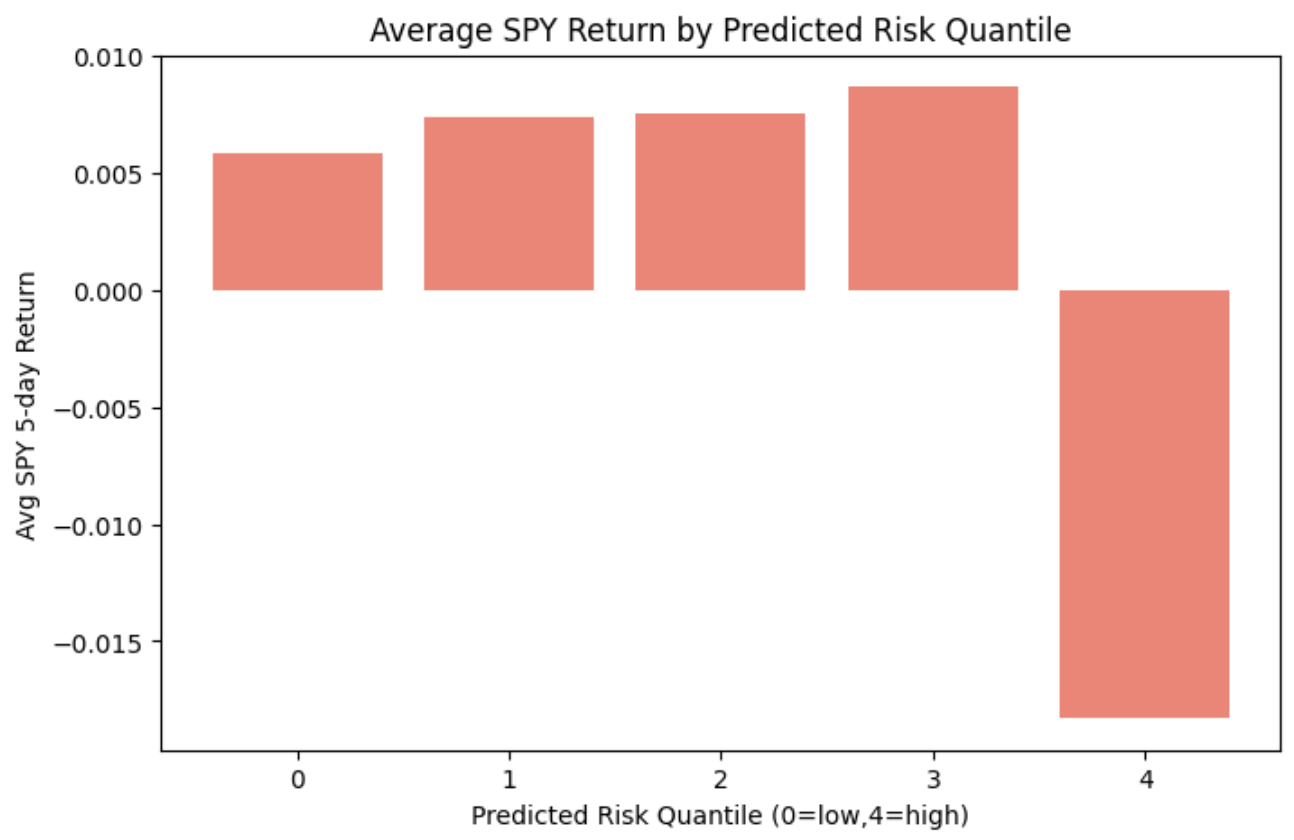}
\caption{Cumulative SPY returns stratified by predicted crash probability quintiles. High predicted risk weeks (Q5) tend to coincide with drawdowns, validating model utility for risk-adjusted positioning.}
\label{fig:risk_quantile_plot}
\end{figure}

\begin{itemize}
\item \textbf{Monotonic risk-return relationship:} There is a clear and economically meaningful monotonic trend: weeks assigned higher predicted crash probability (Q5) systematically coincide with realized drawdowns, whereas the lowest-risk weeks (Q1) experience mostly positive or muted returns. This confirms that the model captures true conditional risk, not just statistical artifacts.

\item \textbf{Practical tactical implications:} The pronounced divergence between extreme quintiles underscores the model’s potential for tactical allocation or protective hedging. A risk-aware investor could reduce equity exposure or hedge during Q5 weeks, while maintaining or even leveraging exposure during Q1–Q2 weeks, improving risk-adjusted performance without sacrificing upside participation.
\end{itemize}

Examining the distribution of cumulative returns within each quintile, Figure~\ref{fig:dist} reveals that the tail risk is concentrated in the highest-risk quintile (HIGH-risk period), while lower-risk quintiles Q1-Q4 (collectively denoted as LOW-risk period) exhibit tight, near-zero variance with modest positive skew. This asymmetry confirms that the model isolates extreme downside events while maintaining stable performance in normal market conditions, a key feature for systematic, risk-managed strategies. To quantify this difference even further: the average 5-day SPY return during HIGH-risk periods comes out to $-1.89\%$, while the average 5-day SPY return during LOW-risk periods comes out to $+0.75\%$.

\subsection{Discussion of Results}

The results validate both the ensemble methodology and the cross-asset, multi-horizon feature engineering pipeline. The combination of MLPs for temporal dependencies and tree-based learners for structured, non-linear interactions allows the model to detect subtle early warning signals not apparent from equity indices alone. SHAP analysis further elucidates causal patterns, indicating that macro and cross-asset stress indicators often precede equity drawdowns, even amid periods of strong headline index momentum driven by factors such as AI hype. Risk quantile performance underscores the potential for systematic alpha generation and risk mitigation through informed position sizing or hedging strategies.

\section{Trading Strategy}
\label{sec:trading_strategy}

Building on the predictive model described in the previous sections, we implement a straightforward long/short trading strategy based on the probability of SPY drawdowns predicted by our ensemble model. Specifically, the strategy operates as follows:

\begin{itemize}
    \item \textbf{Long Position:} Enter a long position in SPY when the predicted probability of a crash in the following week is below a pre-defined threshold of $0.5$ (i.e., the model forecasts a low-risk environment).
    \item \textbf{Short Position:} Enter a short position when the predicted probability of a crash exceeds the threshold of $0.5$, indicating elevated risk.
    \item \textbf{Position Sizing:} Positions are scaled according to the model’s predicted probability, creating a form of risk-adjusted allocation that increases exposure when confidence is high and decreases when uncertainty is elevated.
\end{itemize}

\subsection{Strategy Performance Visualization}
\begin{figure}[H]
    \centering
    \includegraphics[width=\textwidth]{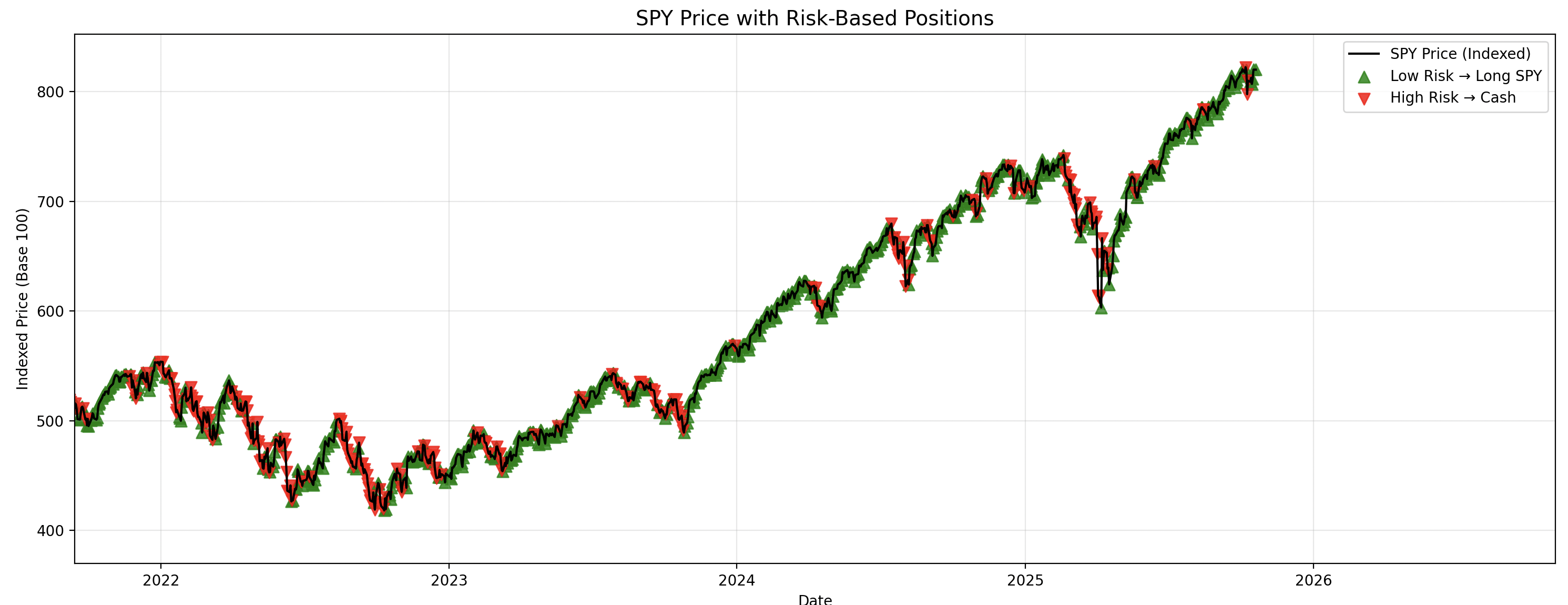}
    \caption{Predicted drawdown probability signal generated by the ensemble model.}
    \label{fig:signal}
\end{figure}

Figure~\ref{fig:signal} shows the evolution of the trading signal alongside realized SPY returns from 2021 to 2025, where green triangles indicate longs and red triangles indicate shorts based on 5-day risk prediction score. These trades are made daily. 

\begin{figure}[H]
    \centering
    \includegraphics[width=\textwidth]{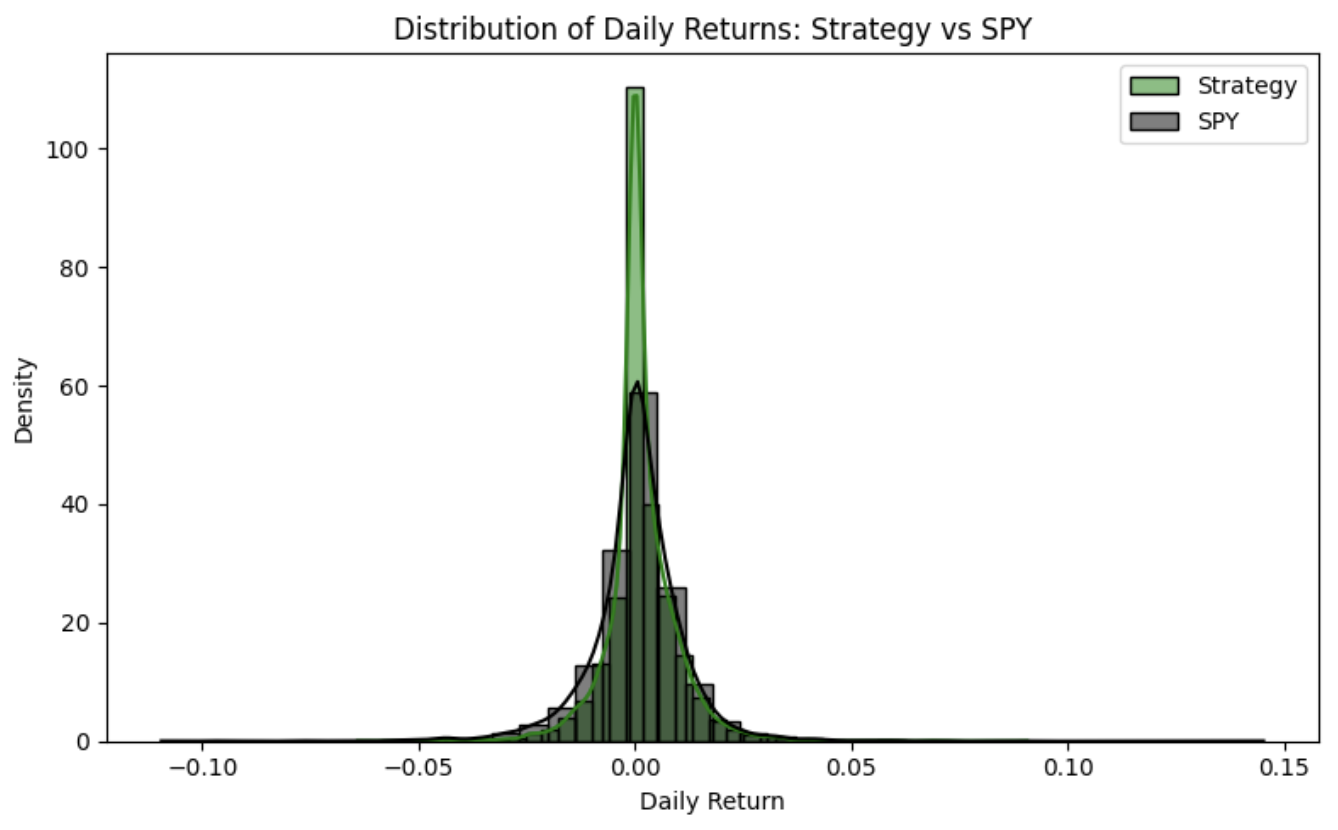}
    \caption{Overlay of SPY daily returns vs. Strategy daily returns}
    \label{fig:strategy}
\end{figure}

Figure~\ref{fig:strategy} presents the distribution of daily returns from the long/short strategy relative to the SPY benchmark over the trained period. Several economically meaningful patterns emerge:

\begin{itemize}
    \item \textbf{Downside protection:} The strategy exhibits a noticeably skinnier left tail compared with SPY's left-side distribution, reflecting its directional accuracy in identifying forward crash weeks and reducing exposure/going short. This indicates effective capture of market stress episodes, limiting extreme losses and highlighting the model’s utility as a conditional crash-protection signal.
    \item \textbf{Upside participation:} The right tail of the strategy closely mirrors the positive return days of SPY, suggesting that the model preserves exposure during normal or bullish periods. In other words, the strategy does not sacrifice upside during calm markets, maintaining participation in standard equity rallies.
    \item \textbf{Central peak near zero:} The pronounced density at near-zero returns corresponds to days when the model adopts a neutral stance. This arises either when predicted crash probabilities are near the decision threshold or when market movements are small, yielding minimal P\&L. Economically, this represents the strategy’s “resting state,” where it neither risks capital on marginal signals nor overreacts to minor market fluctuations.
\end{itemize}
Overall, the distribution highlights the strategy’s asymmetric performance: it systematically mitigates losses during market stress while retaining gains in normal conditions, achieving a balance between downside protection and upside participation—a hallmark of a robust predictive risk-managed trading approach.

\subsection{Strategy Performance Metrics}

Table~\ref{tab:strategy_metrics} summarizes key performance metrics for the long/short strategy constructed from the ensemble predictions over the full dataset. Metrics include risk-adjusted returns, CAPM parameters, and realized SPY returns during high- and low-risk periods identified by the model. 

\begin{table}[H]
\centering
\caption{In-sample trading strategy performance metrics (2005-2025).}
\label{tab:strategy_metrics}
\begin{tabular}{l c}
\toprule
\textbf{Metric} & \textbf{Value} \\
\midrule
Sharpe Ratio & 2.51 \\
Information Ratio vs SPY & 1.73 \\
Maximum Drawdown & -18.12\% \\
Annualized Return & 40.84\% \\
Annualized Volatility & 13.23\% \\
CAPM Alpha (daily) & 0.00111 \\
CAPM Beta & 0.51 \\
T-stat Alpha & 14.03 \\
\bottomrule
\end{tabular}
\end{table}

These in-sample metrics indicate strong conditional performance: the strategy effectively differentiates high versus low crash probability weeks. The high Sharpe and Information Ratios reflect substantial risk-adjusted returns in periods where the model correctly identifies elevated drawdown risk. The moderate beta (0.51) indicates partial decoupling from SPY, consistent with a hedged, long/short approach. Positive daily alpha and highly significant t-statistics further underscore that the ensemble captures information beyond simple market exposure.  

\textit{Note}: These metrics \textbf{do not currently account} for transaction costs nor have they been forward tested. This remains an active consideration for future work.

\subsection{Discussion}

Overall, the strategy underscores the predictive utility of cross-asset, Hurst-based features for short-term drawdown risk management. The exceptionally high ROC-AUC confirms strong model discrimination between crash and non-crash periods, while the performance metrics demonstrate both attractive risk-adjusted returns and effective mitigation of downside exposure. Notably:

\begin{itemize}
    \item \textbf{Risk-Aware Returns:} High annualized return combined with moderate volatility results in a strong Sharpe ratio (2.51).
    \item \textbf{Market Hedging:} CAPM Beta below 1 and high information ratio indicate reduced exposure to overall SPY fluctuations while capturing alpha.
    \item \textbf{Behavior During Stress:} The average SPY return during high-risk periods is negative, validating the model’s early warning capability.
\end{itemize}

\section{Conclusion and Future Work}

In this study, we developed a hybrid machine learning ensemble to predict probabilities of $> 1\%$ weekly SPY drawdowns over the $2005-2025$ period. By integrating tree-based voting models with neural networks, and leveraging cross-asset signals, we achieved strong in-sample classification performance. Feature attribution via SHAP revealed that short-term, conditional drivers—such as SPY and IWM returns, interest rate sensitivities, and commodity tail-risk measures—dominate crash predictions, whereas FX market indicators contribute stabilizing information during calm periods. The divergence between mutual information and SHAP analyses highlights the importance of considering conditional, cross-feature effects when interpreting predictive models in financial markets.

Translating predicted crash probabilities into actionable signals, the strategy demonstrates significant differentiation between high- and low-risk weeks, with elevated risk-adjusted returns, partial decoupling from SPY, and robust conditional alpha. The monotonic relationship between predicted risk and realized returns underscores the practical utility of the model for tactical positioning and protective hedging, even when accounting for market regime dynamics over two decades.

Future work will focus on several extensions: (i) forward-testing and out-of-sample validation to assess robustness in live market conditions; (ii) incorporation of transaction costs, slippage, and liquidity constraints to evaluate real-world feasibility; (iii) exploration of alternative feature representations, including higher-frequency and alternative asset classes; and (iv) deeper causal analysis of cross-asset dynamics to improve interpretability and uncover additional structural market relationships. These directions aim to further enhance both predictive performance and economic interpretability, bridging the gap between machine learning insights and actionable risk management in equity markets.



\begin{thebibliography}{20}

\bibitem{lgbm} Lamoureux, C. G., \& Lastrapes, W. D. (1990). "Persistence in Variance, Structural Change, and the GARCH Model," \emph{Journal of Business \& Economic Statistics}, 8(2), 225–234.

\bibitem{garch} Wei, J., Yang, S., \& Cui, Z. (2025). Integrated GARCH-GRU in Financial Volatility Forecasting. School of Business, Stevens Institute of Technology, Hoboken, NJ, United States. \url{https://arxiv.org/abs/2504.09380v1}

\bibitem{uncertainty} Bekaert, G., Hoerova, M., \& Lo Duca, M. (2013). Risk, uncertainty and monetary policy. \emph{Journal of Monetary Economics}, 60(7), 771–788.

\bibitem{factors} Fama, E. F., \& French, K. R. (1993). Common risk factors in the returns on stocks and bonds. \emph{Journal of Financial Economics}, 33(1), 3–56.

\bibitem{assets} Cont, R. (2001). Empirical properties of asset returns: Stylized facts and statistical issues. \emph{Quantitative Finance}, 1(2), 223–236.

\bibitem{ml} Gu, S., Kelly, B., \& Xiu, D. (2020). Empirical Asset Pricing via Machine Learning. \emph{Review of Financial Studies}, 33(5), 2223–2273.

\bibitem{trees} Krauss, C., Do, X. A., \& Huck, N. (2017). Deep neural networks, gradient-boosted trees, random forests: Statistical arbitrage on the S\&P 500. \emph{European Journal of Operational Research}, 259(2), 689–702.

\bibitem{mi} Cover, T. M., \& Thomas, J. A. (2006). \emph{Elements of Information Theory}. Wiley.

\bibitem{shap} Lundberg, S. M., \& Lee, S.-I. (2017). A unified approach to interpreting model predictions. In \emph{Advances in Neural Information Processing Systems (NeurIPS)}.

\bibitem{rf} Breiman, L. (2001). Random Forests. \emph{Machine Learning}, 45(1), 5–32.

\bibitem{dl} Goodfellow, I., Bengio, Y., \& Courville, A. (2016). \emph{Deep Learning}. MIT Press.

\end{thebibliography}
\end{document}